\documentclass[a4paper,eepic,11pt]{article}
\usepackage{amssymb}
\usepackage{graphicx}
\usepackage{amsmath}

\newtheorem{theorem}{Theorem}

\newtheorem{corollary}[theorem]{Corollary}

\newtheorem{lemma}[theorem]{Lemma}

\newenvironment{proof}[1][Proof]{\textbf{#1.} }{\ \rule{0.5em}{0.5em}}

\begin{document}

\title{Hybrid Rounding Techniques for Knapsack Problems\thanks{%
Supported by the \textquotedblleft Metaheuristics Network\textquotedblright
, grant HPRN-CT-1999-00106, by Swiss National Science Foundation project
20-63733.00/1, \textquotedblleft Resource Allocation and Scheduling in
Flexible Manufacturing Systems\textquotedblright , and by SNF grant
2000-61847.00 to J\"{u}rgen Schmidhuber.}}
\author{Monaldo Mastrolilli and Marcus Hutter \\
{\small IDSIA, Galleria 2, 6928 Manno, Switzerland,} {\small \{monaldo,
marcus\}@idsia.ch}}
\maketitle

\begin{abstract}
We address the classical knapsack problem and a variant in which an upper
bound is imposed on the number of items that can be selected. We show that
appropriate combinations of rounding techniques yield novel and powerful
ways of rounding. As an application of these techniques, we present a
linear-storage Polynomial Time Approximation Scheme (PTAS) and a Fully
Polynomial Time Approximation Scheme (FPTAS) that compute an approximate
solution, of any fixed accuracy, in linear time. This linear complexity
bound gives a substantial improvement of the best previously known
polynomial bounds \cite{Caprara:1998:NFP}.
\end{abstract}

\section{Introduction}

In the classical \textit{Knapsack Problem} (KP) we have a set $N:=\{1,\ldots
,n\}$ of items and a knapsack of limited capacity. To each item we associate
a positive profit $p_{j}$ and a positive weight $w_{j}$. The problem calls
for selecting the set of items with maximum overall profit among those whose
total weight does not exceed the knapsack capacity $c>0$. KP has the
following Integer Linear Programming (ILP) formulation:

\begin{gather}
\text{maximize \ \ \ }\sum_{j\in N}p_{j}x_{j}\text{ \ \ \ \ \ \ \ }
\label{eq:ILP1} \\
\text{subject to \ \ }\sum_{j\in N}w_{j}x_{j}\leq c  \label{eq:ILP2} \\
\text{ \ \ \ \ \ \ \ \ \ \ \ \ \ \ \ \ \ \ \ \ \ \ \ \ \ }x_{j}\in \{0,1\},%
\text{ \ \ }j\in N,  \label{eq:ILP3}
\end{gather}%
where each binary variable $x_{j}$, $j\in N$, is equal to 1 if and only if
item $j$ is selected. In general, we cannot take all items because the total
weight of the chosen items cannot exceed the knapsack capacity $c$. In the
sequel, without loss of generality, we assume that $\sum_{j\in N}w_{j}>c$
and $w_{j}\leq c$ for every $j\in N$.

The \textit{k-item Knapsack Problem} (kKP), is a KP in which an upper bound
of $k$ is imposed on the number of items that can be selected in a solution.
The problem can be formulated as (\ref{eq:ILP1})-(\ref{eq:ILP3}) with the
additional constraint

\begin{equation}
\sum_{j\in N}x_{j}\leq k,  \label{eq:ILP4}
\end{equation}
with $1\leq k\leq n$.

KP has widely been studied in the literature, see the book of Martello and
Toth \cite{MarToth90} for a comprehensive illustration of the problem. kKP
is the subproblem to be solved when instances of the Cutting Stock Problem
with cardinality constraints are tackled by column generation techniques.
kKP also appears in processor scheduling problems on computers with $k$
processors and shared memory. Furthermore, kKP could replace KP in the
separation of cover inequalities, as outlined in \cite{Caprara:1998:NFP}.

Throughout our paper let $OPT$ denote the optimal solution value to the
given instance and $w(F)=\sum_{j\in F}w_{j}$ and $p(F)=\sum_{j\in F}p_{j}$,
where $F\subseteq N$. An algorithm $A$ with solution value $z^{A}$ is called
a $(1-\varepsilon )$-\textit{approximation algorithm}, $\varepsilon \in
(0,1) $, if $z^{A}\geq (1-\varepsilon )OPT$ holds for all problem instances.
We will also call $\varepsilon $ the \textit{performance ratio} of $A$.

\paragraph{Known Results}

It is well known that KP is NP-hard but pseudopolynomially solvable through
dynamic programming, and the same properties hold for kKP \cite%
{Caprara:1998:NFP}. Basically, the developed approximation approaches for KP
and kKP can be divided into three groups:

\begin{enumerate}
\item \textit{Approximation algorithms}. For KP the classical $\frac{1}{2}$%
-approximation algorithm (see e.g. \cite{app:Lawler:77}) needs only $O(n)$
running time. A performance ratio of $\frac{1}{2}$ can be obtained also for
kKP by rounding the solution of the linear programming relaxation of the
problem (see \cite{Caprara:1998:NFP}); this algorithm can be implemented to
run in linear time when the LP relaxation of kKP is solved by using the
method by Megiddo and Tamir \cite{MegTam:93}.

\item \textit{Polynomial time approximation schemes} (PTAS) reach any given
performance ratio and have a running time polynomial in the length of the
encoded input. The best schemes currently known requiring linear space are
given in Caprara et al. \cite{Caprara:1998:NFP}: they yield a performance
ratio of $\varepsilon $ within $O(n^{\left\lceil 1/\varepsilon \right\rceil
-2}+n\log n)$ and $O(n^{\left\lceil 1/\varepsilon \right\rceil -1})$ running
time, for KP and kKP respectively.

\item \textit{Fully polynomial time approximation schemes} (FPTAS) also
reach any given performance ratio and have a running time polynomial in the
length of the encoded input and in the reciprocal of the performance ratio.
This improvement compared to 1. and 2. is usually paid off by larger space
requirements, which increases rapidly with the accuracy $\varepsilon $. The
first FPTAS for KP was proposed by Ibarra and Kim \cite{IbarraKim:1975},
later on improved by Lawler \cite{app:Lawler:77} and Kellerer and Pferschy
\cite{Kellerer:1998:NFP}. In Caprara et al. \cite{Caprara:1998:NFP} it is
shown that kKP admits an FPTAS that runs in $O(nk^{2}/\varepsilon )$ time.
\end{enumerate}

\paragraph{New Results}

Rounding the input is a widely used technique to obtain polynomial time
approximation schemes \cite{H95}. Among the developed rounding techniques,
arithmetic or geometric rounding are the most successfully and broadly used
ways of rounding to obtain a simpler instance that may be solved in
polynomial time (see Sections \ref{Sect:arithmetic rounding} and \ref%
{Sect:Geometric rounding} for an application of these techniques to kKP). We
contribute by presenting a new technical idea. We show that appropriate
combinations of arithmetic and geometric rounding techniques yield novel and
powerful rounding methods. To the best of our knowledge, these techniques
have never been combined together. By using the described rounding
techniques, we present a PTAS for kKP requiring linear space and running
time $O(n+k\cdot (1/\varepsilon )^{O(1/\varepsilon )})$. Our algorithm is
clearly superior to the one in \cite{Caprara:1998:NFP}, and it is worth
noting that the running time contains no exponent on $n$ dependent on $%
\varepsilon $. Since KP is a special case of kKP, we also speed up the
previous result for KP to $O(n\cdot (1/\varepsilon )^{O(1/\varepsilon )})$.
Finally we present a faster FPTAS for kKP that runs in $O(n+k/\varepsilon
^{4}+1/\varepsilon ^{5})$ time and has a bound of $O(n+1/\varepsilon ^{4})$
on space requirements.

\section{Rounding techniques for kKP\label{Sect:reduced N}}

The aim of this section is to transform any input into one with a smaller
size and a simpler structure without dramatically decreasing the objective
value. We discuss several transformations of the input problem. Some
transformations may potentially decrease the objective function value by a
factor of $1-O(\varepsilon )$, so we can perform a constant number of them
while still staying within $1-O(\varepsilon )$ of the original optimum.
Others are transformations which do not decrease the objective function
value. When we describe the first type of transformation, we shall say it
produces $1-O(\varepsilon )$ \textit{loss}, while the second produces
\textit{no} \textit{loss}.

Let $P^{H}$ denote the solution value obtained in $O(n)$ time by employing
the $1/2$-approximation algorithm $H^{\frac{1}{2}}$ for kKP described in
\cite{Caprara:1998:NFP}. In \cite{Caprara:1998:NFP}, it is shown that
\begin{equation}
2P^{H}\geq P^{H}+p_{\max }\geq OPT\geq P^{H},  \label{Eq:ineq}
\end{equation}
where $p_{\max }=\max_{j}p_{j}$.

Throughout this section we restrict our attention to feasible solutions with
at most $\gamma $ items, where $\gamma $ is a positive integer not greater
than $k$. The first observation is that at most an $\varepsilon $-fraction
of the optimal profit $OPT$ is lost by discarding all items $j$ where $%
p_{j}\leq \varepsilon P^{H}/\gamma $, since at most $\gamma $ items can be
selected and $P^{H}\leq OPT$. From now on, consider the reduced set of items
with profit values greater than $\varepsilon P^{H}/\gamma $, with $%
1-\varepsilon $ loss.

In order to reduce further the set of items, a useful insight is that when
profits are identical we pick items in non-decreasing order of weight. Since
the optimal profit is at most $2P^{H}$, for each fixed $\bar{p}\in \left\{
p_{1},...,p_{n}\right\} $, we can keep the first $\bar{n}=\min \left\{
\gamma ,\left\lfloor \frac{2P^{H}}{\bar{p}}\right\rfloor \right\} $ items
with the smallest weights, and discard the others with no loss. Of course,
we cannot hope to obtain a smaller instance if all profits are different. In
the following, we show how the number of different profits can be reduced by
rounding the original profits. We revise two rounding techniques and show
that an appropriate combination of both yields to a better result. We call a
profit value $\bar{p}$ \textit{large} if $\bar{p}>\frac{2P^{H}}{\gamma }$,
and \textit{small} otherwise.

\subsection{Arithmetic rounding\label{Sect:arithmetic rounding}}

A sequence $a_{1},a_{2},...$ is called an \textit{arithmetic sequence} if,
and only if, there is a constant $d$ such that $a_{i}=a_{1}+d\cdot (i-1)$,
for all integers $i\geq 1$. Let us consider the arithmetic sequence $%
S_{a}(\gamma )$ obtained by setting $a_{1}=d=\varepsilon P^{H}/\gamma $. We
transform the given instance into a more structured one by rounding each
profit down to the nearest value in $S_{a}(\gamma )$. Since in the rounded
instance the profit of each item is decreased by at most $\varepsilon
P^{H}/\gamma $, and at most $\gamma $ items can be selected, the solution
value of the transformed instance potentially decreases by $\varepsilon
P^{H} $. Of course, by restoring the original profits we cannot decrease the
objective function value, and therefore, with $1-\varepsilon $ loss, we can
assume that every possible profit is equal to $a_{i}=\frac{\varepsilon P^{H}%
}{\gamma }\cdot i$ for some $i\geq 1$. Furthermore, since $p_{\max
}=\max_{j\in N}p_{j}\leq P^{H}$, the number of different profits is now
bounded by $\left\lfloor \frac{\gamma p_{\max }}{\varepsilon P^{H}}%
\right\rfloor \leq \left\lfloor \frac{\gamma }{\varepsilon }\right\rfloor $.

The largest number $n_{i}$ of items with profit $a_{i}$, for $%
i=1,...,\left\lfloor \frac{\gamma }{\varepsilon }\right\rfloor $, that can
be involved in any feasible solution is bounded by
\begin{equation*}
n_{i}\leq \min \{\gamma ,\left\lfloor \frac{OPT}{\frac{\varepsilon P^{H}}{%
\gamma }i}\right\rfloor \}\leq \min \{\gamma ,\left\lfloor \frac{2\gamma }{%
\varepsilon i}\right\rfloor \},
\end{equation*}
and we can keep the first $n_{i}$ items with the smallest weights, and
discard the others with no loss. Now, the number of items with profit $a_{i}$
is at most $\gamma $, if $a_{i}$ is a small profit (i.e. when $%
i=1,...,\left\lfloor \frac{2}{\varepsilon }\right\rfloor $), and at most $%
\left\lfloor \frac{2\gamma }{\varepsilon i}\right\rfloor $ otherwise ($%
i=\left\lfloor \frac{2}{\varepsilon }\right\rfloor +1,...,\left\lfloor \frac{%
\gamma }{\varepsilon }\right\rfloor $). Thus, by applying the described
arithmetic rounding we have at most $\left\lfloor 2/\varepsilon
\right\rfloor \gamma $ items with small profits and $\allowbreak
\sum_{i=\left\lfloor \frac{2}{\varepsilon }\right\rfloor +1}^{\left\lfloor
\frac{\gamma }{\varepsilon }\right\rfloor }\left\lfloor \frac{2\gamma }{%
\varepsilon i}\right\rfloor $ with large profits. Recall that when a
summation can be expressed as $\sum_{k=x}^{y}f(k)$, where $f(k)$ is a
monotonically decreasing function, we can approximate it by an integral
(see, e.g. \cite{CLR92} p. 50): $\int_{x}^{y+1}f(k)dk\leq
\sum_{k=x}^{y}f(k)\leq \int_{x-1}^{y}f(k)dk$. Furthermore, we are assuming
that $0<\varepsilon <1$, and recall that $\ln (1+x)\geq x/(1+x)$, for $x>-1$%
. Therefore, the total number of items in the transformed instance is
bounded by
\begin{eqnarray*}
\left\lfloor \frac{2}{\varepsilon }\right\rfloor \gamma
+\sum_{i=\left\lfloor \frac{2}{\varepsilon }\right\rfloor +1}^{\left\lfloor
\frac{\gamma }{\varepsilon }\right\rfloor }\left\lfloor \frac{2\gamma }{%
\varepsilon i}\right\rfloor &\leq &\frac{2}{\varepsilon }\gamma +\frac{2}{%
\varepsilon }\gamma \sum_{i=\left\lfloor \frac{2}{\varepsilon }\right\rfloor
+1}^{\left\lfloor \frac{\gamma }{\varepsilon }\right\rfloor }\frac{1}{i} \\
&\leq &\frac{2\gamma }{\varepsilon }(1+\int_{\frac{2}{\varepsilon }-1}^{%
\frac{\gamma }{\varepsilon }}\frac{di}{i})=\allowbreak \frac{2\gamma }{%
\varepsilon }(1+\ln \gamma -\ln \left( 2-\varepsilon \right) ) \\
&\leq &\frac{2\gamma }{\varepsilon }(1+\ln \gamma )=O(\frac{\gamma }{%
\varepsilon }\ln \gamma ).
\end{eqnarray*}
We see that by applying the described arithmetic rounding we have at most $%
2\gamma /\varepsilon $ items with small profits and $\allowbreak \frac{%
2\gamma }{\varepsilon }\ln \gamma $ with large profits.

A natural question is to see if the provided bound is tight. Consider an
instance having $\gamma $ items for each distinct profit $\frac{\varepsilon
P^{H}}{\gamma }\cdot i$, where $i=1,...,\left\lfloor \frac{\gamma }{%
\varepsilon }\right\rfloor $. Observe that by applying the described
arithmetic rounding, we have exactly $\left\lfloor 2/\varepsilon
\right\rfloor \gamma $ items with small profits and $\allowbreak
\sum_{i=\left\lfloor \frac{2}{\varepsilon }\right\rfloor +1}^{\left\lfloor
\frac{\gamma }{\varepsilon }\right\rfloor }\left\lfloor \frac{2\gamma }{%
\varepsilon i}\right\rfloor $ with large profits. What remains to be shown
is to bound the number of items to be $\Omega (\frac{\gamma }{\varepsilon }%
\ln \gamma )$:

\begin{eqnarray*}
\left\lfloor \frac{2}{\varepsilon }\right\rfloor \gamma
+\sum_{i=\left\lfloor \frac{2}{\varepsilon }\right\rfloor +1}^{\left\lfloor
\frac{\gamma }{\varepsilon }\right\rfloor }\left\lfloor \frac{2\gamma }{%
\varepsilon i}\right\rfloor &\geq &\frac{2}{\varepsilon }\gamma -1+\sum_{i=%
\frac{2}{\varepsilon }+1}^{\frac{\gamma }{\varepsilon }-1}(\frac{2\gamma }{%
\varepsilon i}-1) \\
&\geq &\frac{2}{\varepsilon }\gamma -1+\frac{2}{\varepsilon }\gamma \sum_{i=%
\frac{2}{\varepsilon }+1}^{\frac{\gamma }{\varepsilon }-1}\frac{1}{i}-(\frac{%
\gamma }{\varepsilon }-2) \\
&\geq &\frac{\gamma }{\varepsilon }(1+2\int_{\frac{2}{\varepsilon }+1}^{%
\frac{\gamma }{\varepsilon }}\frac{di}{i})=\allowbreak \frac{\gamma }{%
\varepsilon }(1+2\ln \gamma -2\ln \left( 2+\varepsilon \right) ) \\
&=&\Omega (\frac{\gamma }{\varepsilon }\ln \gamma ).
\end{eqnarray*}

\subsection{Geometric rounding\label{Sect:Geometric rounding}}

A sequence $a_{1},a_{2},...$ is called a \textit{geometric sequence} if, and
only if, there is a constant $r$ such that $a_{i}=a_{1}\cdot r^{i-1}$, for
all integers $i\geq 1$. Let us consider the geometric sequence $S_{g}(\gamma
)$ obtained by setting $a_{1}=\varepsilon P^{H}/\gamma $ and $r=\frac{1}{%
1-\varepsilon }$. We round each profit down to the nearest value among those
of $S_{g}(\gamma )$. Since $a_{i}=(1-\varepsilon )a_{i+1}$, for $i\geq 1$,
each item profit is at most decreased by a factor\footnote{%
This is true only for profits larger than $a_{1}$; recall that we are
assuming, with $1-\varepsilon $ loss, that all profits are larger than $a_{1}
$.} of $1-\varepsilon $, and consequently, the solution value of the
transformed instance potentially decreases by the same factor of $%
1-\varepsilon $. Therefore, with $1-2\varepsilon $ loss, we can assume that
every possible profit is equal to $a_{i}=\frac{\varepsilon P^{H}}{\gamma }%
\cdot (\frac{1}{1-\varepsilon })^{i-1}$ for some $i\geq 1$. Furthermore,
since $p_{\max }\leq P^{H}$, the number of different profits is bounded by
the biggest integer $\beta $ such that
\begin{equation*}
\frac{\varepsilon P^{H}}{\gamma }\cdot (\frac{1}{1-\varepsilon })^{\beta
-1}\leq P^{H}.
\end{equation*}%
Since $\ln (\frac{1}{1-\varepsilon })\geq \varepsilon $, we have $\beta
-1\leq \frac{\ln (\gamma /\varepsilon )}{\ln (\frac{1}{1-\varepsilon })}\leq
\frac{1}{\varepsilon }\ln \frac{\gamma }{\varepsilon }$. In any feasible
solution, the largest number $n_{i}$ of items with profit $a_{i}$, for $%
i=1,...,\beta $, is bounded by
\begin{equation*}
n_{i}\leq \min \{\gamma ,\left\lfloor \frac{OPT}{\frac{\varepsilon P^{H}}{%
\gamma }\cdot (\frac{1}{1-\varepsilon })^{i-1}}\right\rfloor \}\leq \min
\{\gamma ,\left\lfloor \frac{2\gamma }{\varepsilon }(1-\varepsilon
)^{i-1}\right\rfloor \},
\end{equation*}%
and we can keep the first $n_{i}$ items with the smallest weights, and
discard the others with no loss. Let $\alpha =\left\lfloor \frac{\ln
\left\lfloor 2/\varepsilon \right\rfloor }{\ln \frac{1}{1-\varepsilon }}%
\right\rfloor +1$. Again, the number of items with profit $a_{i}$ is at most
$\gamma $, if $a_{i}$ is a small profit (i.e. when $1\leq i\leq \alpha $),
and at most $\left\lfloor \frac{2\gamma }{\varepsilon }(1-\varepsilon
)^{i-1}\right\rfloor $ otherwise ($i=\alpha +1,...,\beta $). Therefore, the
total number of items in the transformed instance is bounded by
\begin{equation*}
\alpha \gamma +\sum_{i=\alpha +1}^{\beta }\left\lfloor \frac{2\gamma }{%
\varepsilon }(1-\varepsilon )^{i-1}\right\rfloor \leq (\frac{1}{\varepsilon }%
\ln (2/\varepsilon )+1)\gamma +\allowbreak \frac{2\gamma }{\varepsilon }=O(%
\frac{\gamma }{\varepsilon }\ln \frac{1}{\varepsilon }).
\end{equation*}%
Moreover, we can easily show that this bound is tight. Consider an instance
having $\gamma $ items for each distinct profit $\frac{\varepsilon P^{H}}{%
\gamma }\cdot (\frac{1}{1-\varepsilon })^{i-1}$, where $i=1,...,\beta $. By
applying the described geometric rounding technique, we have exactly $\alpha
\gamma $ items with small profits and $\allowbreak \sum_{i=\alpha +1}^{\beta
}\left\lfloor \frac{2\gamma }{\varepsilon }(1-\varepsilon
)^{i-1}\right\rfloor $ with large profits. What remains to be shown is to
bound the number of items to be $\Omega (\frac{\gamma }{\varepsilon }\ln
\frac{1}{\varepsilon })$:
\begin{eqnarray*}
\alpha \gamma +\sum_{i=\alpha +1}^{\beta }\left\lfloor \frac{2\gamma }{%
\varepsilon }(1-\varepsilon )^{i-1}\right\rfloor  &\geq &\alpha \gamma \geq
\frac{\ln \left\lfloor 2/\varepsilon \right\rfloor }{\ln \frac{1}{%
1-\varepsilon }}\gamma  \\
&\geq &\gamma \frac{1-\varepsilon }{\varepsilon }\ln \frac{2-\varepsilon }{%
\varepsilon }=\Omega (\frac{\gamma }{\varepsilon }\ln \frac{1}{\varepsilon }%
).
\end{eqnarray*}

We see that by applying the geometric rounding we have at most $\gamma
/\varepsilon $ items with large profit, while $O(\frac{\gamma }{\varepsilon }%
\ln \frac{1}{\varepsilon })$ items with small profits. Contrary to
arithmetic rounding, the set of items that has been reduced most is the set
with large profits. This suggests us to combine the described rounding
techniques as described in the following subsection.

\subsection{Parallel Arithmetic \& Geometric rounding\label{Sect:arith_geo}}

We use arithmetic rounding for the set of items with small profits and
geometric rounding for large items. Let us say that these two techniques are
applied in \textquotedblleft parallel\textquotedblright . More formally, let
us consider the \textit{hybrid }sequence $S_{ag}(\gamma )=(a_{1},a_{2},...)$
defined by setting
\begin{equation*}
a_{i}=\frac{\varepsilon P^{H}}{\gamma }\cdot \left\{
\begin{array}{l}
i \\
(\frac{1}{1-\varepsilon })^{\alpha +i-\left\lfloor 2/\varepsilon
\right\rfloor -1}%
\end{array}%
\right.
\begin{array}{l}
\text{for }i=1,...,\left\lfloor 2/\varepsilon \right\rfloor \text{,} \\
\text{otherwise.}%
\end{array}%
\end{equation*}%
We round each profit down to the nearest value among those of $S_{ag}(\gamma
)$. Now, consider each set $Z_{i}$ of items with the same rounded profit
value $a_{i}$, and take the first
\begin{equation*}
n_{i}=\left\{
\begin{array}{l}
\gamma \\
\left\lfloor \frac{2\gamma }{\varepsilon }(1-\varepsilon )^{\alpha
+i-\left\lfloor 2/\varepsilon \right\rfloor -1}\right\rfloor%
\end{array}%
\right.
\begin{array}{l}
\text{for }i=1,...,\left\lfloor 2/\varepsilon \right\rfloor \text{,} \\
\text{otherwise,}%
\end{array}%
\end{equation*}%
items with the smallest weights we get in $Z_{i}$. Selecting the first $%
n_{i} $ items with the smallest weights can be done in $O(|Z_{i}|)$ time.
That is, $O(|Z_{i}|)$ time is sufficient to select the $n_{i}$-th item with
the smallest weight (see \cite{median-finding}) and only $O(|Z_{i}|)$
comparisons are needed to extract the $n_{i}-1$ items with smaller weights.
Therefore the amortized time is linear.

By using the arithmetic rounding technique for small items, we have at most $%
2\gamma /\varepsilon $ small items with $1-\varepsilon $ loss (see Section %
\ref{Sect:arithmetic rounding}). While, by using the geometric rounding
technique described in Section \ref{Sect:Geometric rounding} for large
items, we have at most $\gamma /\varepsilon $ large items with $%
1-\varepsilon $ loss. The resulting transformed instance has at most $%
3\gamma /\varepsilon $ items with $1-2\varepsilon $ loss. Furthermore, let $%
\psi =\beta -\alpha +\left\lfloor 2/\varepsilon \right\rfloor +1$. Observe
that the $\psi $-th element of $S_{ag}(\gamma )$ is larger than $P^{H}$,
i.e. $a_{\psi }>P^{H}$. Consider any subset $S\subseteq N$ of items with at
most $\gamma $ items, and let $x_{i}$ denote the total number of items from $%
S$ with profit in interval $[a_{i},a_{i+1})$, $i=1,2,...\psi -1$. Let us
call vector $(x_{1},x_{2},...,x_{\psi -1})$ an $S$\textit{-configuration}.
It is easy to see that by using the reduced set of items it is always
possible to compute a solution having the same $S$-configuration, as any set
$S\subseteq N$ with $\gamma $ items. Summarizing:

\begin{lemma}
\label{Lemma: gamma}For any positive integer $\gamma \leq k$, it is possible
to compute in linear time a reduced set $N_{\gamma }\subseteq N$ of items
with at most $3\gamma /\varepsilon $ items, such that, for any subset $%
S\subseteq N$ with at most $\gamma $ items, there exists a subset $S_{\gamma
}\subseteq N_{\gamma }$ such that $S_{\gamma }$ is the subset of $N$ having
the same configuration as $S$ and with the least weights.
\end{lemma}

\begin{corollary}
\label{Corollary: gamma}For any subset $S\subseteq N$ with at most $\gamma $
items, there exists a subset $S_{\gamma }\subseteq N_{\gamma }$ with $%
w(S_{\gamma })\leq w(S)$, $|S_{\gamma }|=|S|$ and $p(S_{\gamma })\geq
p(S)-2\varepsilon \cdot OPT$.
\end{corollary}

\section{An improved PTAS for kKP}

Our PTAS for kKP improves the scheme of Caprara et al. \cite%
{Caprara:1998:NFP}, and in fact it strongly builds on their ideas. However,
there are several differences where a major one is the use of two reduced
sets of items instead of the entire set $N$: let $\ell :=\min \{\lceil
1/\varepsilon \rceil -2,k\}$, where $\varepsilon \leq 1/2$ is an arbitrary
small rational number; our algorithm uses sets $N_{k}$ and $N_{\ell }$
computed by using the Arithmetic \& Geometric rounding technique (see Lemma %
\ref{Lemma: gamma}) when $\gamma :=k$ and $\gamma :=\ell $, respectively.

For any given instance of kKP, the approximation scheme performs the
following five steps (S-1)-(S-5).

\begin{enumerate}
\item[(S-1)] Initialize the solution $A$ to be the empty set and set the
corresponding value $P^{A}$ to $0$.

\item[(S-2)] Compute the reduced sets $N_{k}$ and $N_{\ell }$.

\item[(S-3)] Compute $P^{H}$, i.e. the solution value returned by $H^{\frac{1%
}{2}}$ when applied to the whole set of instances.

\item[(S-4)] Consider each $L\subseteq N_{\ell }$ such that $|L|\leq \ell $.
If $w(L)\leq c$, consider sequence $S_{ag}(\ell )=(a_{1},a_{2},...)$ and let
$h$ be the smallest integer such that $\min_{j\in L}p_{j}<a_{h+1}$ (assume $%
\min_{j\in L}p_{j}=0$ if $L=\varnothing $). Apply algorithm $H^{\frac{1}{2}}$
to the subinstance $S$ defined by item set $\{i\in N_{k}\backslash
L:p_{i}<a_{h+1}\}$, by capacity $c-w(L)$ and by cardinality upper bound $%
k-\ell $. Let $T$ and $P^{H}(S)$ denote the solution and the solution value
returned by $H^{\frac{1}{2}}$ when applied to $S$, respectively. If $%
p(L)+P^{H}(S)>P^{A}$ let $A:=L\cup T$ and $P^{A}:=p(L)+P^{H}(S)$.

\item[(S-5)] Return solution $A$ of value $P^{A}$.
\end{enumerate}

Observe that in step (S-4), subsets $L$ are computed by considering just the
items from $N_{\ell }$. On the other hand, in step (S-4), we remark that the
subinstances $S$ are defined by using items from $N_{k}$.

\subsection{Analysis of the Algorithm}

Step (S-2) can be performed in $O(n)$ time by Lemma \ref{Lemma: gamma}. Step
(S-3) runs in $O(n)$ time \cite{Caprara:1998:NFP}. In step (S-4) the
algorithm considers $O(|N_{\ell }|+|N_{\ell }|^{2}+...+|N_{\ell }|^{\ell
})=O(|N_{\ell }|^{\ell })$ subsets. For each $L$ the definition of
subinstance $S$ requires $O(|N_{k}|\cdot \ell )$ time. Algorithm $H^{\frac{1%
}{2}}$ applied to subinstance $S$ runs in $O(|S|)=O(|N_{k}|)$ time \cite%
{Caprara:1998:NFP}. By Lemma \ref{Lemma: gamma}, $|N_{k}|=O(k/\varepsilon )$
and $|N_{\ell }|=O(\ell /\varepsilon )$. Therefore, step (S-4) is performed
in $O(|N_{\ell }|^{\ell }\cdot |N_{k}|\cdot \ell )=O(k\cdot (\frac{\ell }{%
\varepsilon })^{\ell +1})=k\cdot (1/\varepsilon )^{O(1/\varepsilon )}$. It
follows that the overall running time of the algorithm is $O(n+k\cdot
(1/\varepsilon )^{O(1/\varepsilon )})$, and it is not difficult to check
that steps (S-1)-(S-5) require linear space. What remains to be shown is
that steps (S-1)-(S-5) return a $(1-O(\varepsilon ))$-approximate solution.

Consider sequence $S_{ag}(\ell )=(a_{1},a_{2},...)$. Let $\left\{
j_{1},...,j_{\ell },...\right\} $ be the set of items in an optimal solution
ordered so that $p_{j_{1}}\geq ...\geq p_{j_{\ell }}\geq ...$, and let $%
\lambda \leq \ell $ be the largest integer such that $p_{j_{\lambda }}\geq
a_{1}$ (if there is no item in the optimal solution with profit $\geq a_{1}$
then set $\lambda =0$). Let $L^{\ast }=\left\{ j_{1},...,j_{\lambda
}\right\} $ be the subset (possibly empty) obtained by picking the first $%
\lambda $ items with the largest profits. Consider subinstance $S^{\ast }$
defined by item set
\begin{equation*}
I_{S^{\ast }}=\{i\in N\backslash L^{\ast }\left\vert
\begin{array}{l}
p_{i}\leq \min_{j\in L^{\ast }}p_{j}\text{ if }L^{\ast }\neq \varnothing \\
p_{i}<a_{1}\text{, \ otherwise.}%
\end{array}%
\right\} ,
\end{equation*}%
by capacity $c-w(L^{\ast })$ and by cardinality upper bound $k-\lambda $.
Clearly,
\begin{equation}
p(L^{\ast })+OPT_{S^{\ast }}=OPT,  \label{Eq:ineq0}
\end{equation}%
where $OPT_{S^{\ast }}$ denotes the optimal value of instance $S^{\ast }$.
Now, consider the reduced set $N_{k}$ and subinstance $S_{k}^{\ast }$
defined by item set
\begin{equation}
I_{S_{k}^{\ast }}=\{i\in N_{k}\backslash L^{\ast }\left\vert
\begin{array}{l}
p_{i}\leq \min_{j\in L^{\ast }}p_{j}\text{ if }L^{\ast }\neq \varnothing \\
p_{i}<a_{1}\text{, \ otherwise.}%
\end{array}%
\right\} ,  \label{Def:Isk}
\end{equation}%
by capacity $c-w(L^{\ast })$ and by cardinality upper bound $k-\lambda $. By
Corollary \ref{Corollary: gamma}, we have
\begin{equation}
OPT_{S_{k}^{\ast }}\geq OPT_{S^{\ast }}-2\varepsilon OPT,  \label{Eq:ineq1}
\end{equation}%
where $OPT_{S_{k}^{\ast }}$ denotes the optimal value of instance $%
S_{k}^{\ast }$.

Let us use $L$ to denote the set of items having the same configuration as $%
L^{\ast }$ and the least weights. By Lemma \ref{Lemma: gamma}, in one of the
iterations of step (S-4), set $L$ is considered. In the remainder, let us
focus on this set $L$, and consider the corresponding subinstance $S$
defined in step (S-4). By Corollary \ref{Corollary: gamma}, we have
\begin{equation}
p(L)\geq p(L^{\ast })-2\varepsilon OPT.  \label{Eq:ineq2}
\end{equation}%
We need to show that the optimal solution value $OPT_{S}$ of instance $S$
cannot be smaller than $OPT_{S_{k}^{\ast }}$.

\begin{lemma}
\label{Lemma:Z}$OPT_{S}\geq $ $OPT_{S_{k}^{\ast }}$.
\end{lemma}

\begin{proof}
Assume $L$ having the same configuration as $L^{\ast }$. Recall that the
subinstance $S$ is defined by item set $I_{S}=\{i\in N_{k}\backslash
L:p_{i}<a_{h+1}\}$, where $a_{h+1}$ is the term of sequence $S_{ag}(\ell
)=(a_{1},a_{2},...)$ such that $h$ is the smallest integer with $\min_{j\in
L}p_{j}<a_{h+1}$ (see step (S-4)). On the other hand, the subinstance $%
S_{k}^{\ast }$ is defined by item set $I_{S_{k}^{\ast }}$ (see (\ref{Def:Isk}%
)). If $L^{\ast }=\varnothing $ then $S=S_{k}^{\ast }$ and the claim follows.

Otherwise ($L^{\ast }\neq \varnothing $), since we are assuming that $L$ has
the same configuration as $L^{\ast }$, there are no items from $L^{\ast }$
with profit in intervals $[a_{i},a_{i+1})$, for $i<h$. Therefore, we have $%
\min_{j\in L^{\ast }}p_{j}\geq a_{h}$ and $\{i\in N_{k}\backslash
L:p_{i}\leq a_{h}\}=\{i\in N_{k}\backslash L^{\ast }:p_{i}\leq a_{h}\}$.
Furthermore, since there is at least one item from $L$ with profit in
interval $[a_{h},a_{h+1})$ (recall we are assuming $L^{\ast }\neq
\varnothing $), and since $L^{\ast }$ has the same configuration as $L$,
there exists an item from $L^{\ast }$ with profit $p_{j}<$ $a_{h+1}$ and,
therefore, $\min_{j\in L^{\ast }}p_{j}<a_{h+1}$. It follows that $%
I_{S_{k}^{\ast }}\subseteq \{i\in N_{k}\backslash L^{\ast }:p_{i}<a_{h+1}\}$.

By the previous arguments, the items of $S_{k}^{\ast }$, except those
belonging to $A_{h}=\{i\in N_{k}\cap L:a_{h}\leq p_{i}<a_{h+1}\}$, are also
items of $S$, i.e.,
\begin{equation*}
I_{S_{k}^{\ast }}\subseteq I_{S}\cup A_{h}.
\end{equation*}

If there exists an optimal solution for $S_{k}^{\ast }$ such that no one of
the items from $A_{h}$ is selected, then $OPT_{S}\geq $ $OPT_{S_{k}^{\ast }}$%
, since the knapsack capacity of $S_{k}^{\ast }$ is not greater than the one
of $S$, i.e. $c-w(L)\geq c-w(L^{\ast })$ (recall that $L$ is the subset
having the same configuration as $L^{\ast }$ with the smallest weights).

Otherwise, let $G_{1}$ be the subset of items from $A_{h}$ in an optimal
solution for $S_{k}^{\ast }$, and let $g:=|G_{1}|$. Let $G_{2}$ be any
subset of $\{i\in L^{\ast }\backslash L:a_{h}\leq p_{i}<a_{h+1}\}$
containing exactly $g$ items. It is easy to see that $G_{2}$ exists (recall
that $L$ and $L^{\ast }$ have the same configurations and $A_{h}\subseteq L$%
). Furthermore, since $G_{2}\subseteq L^{\ast }$ and $G_{1}\subseteq
I_{S_{k}^{\ast }}$, we have
\begin{equation}
\min_{j\in G_{2}}p_{j}\geq \max_{j\in G_{1}}p_{j}.  \label{eq:G1G2}
\end{equation}%
Observe that $w(L^{\ast })-w(L)\geq w(G_{2})-w(G_{1})$. Therefore, the
knapsack capacity $c-w(L)$ of $S$ cannot be smaller than $c-w(L^{\ast
})+w(G_{2})-w(G_{1})$. The solution $G_{12}$ obtained from the optimal
solution for $S_{k}^{\ast }$ by replacing the items from $G_{1}$ with those
from $G_{2}$, requires a knapsack of capacity bounded by $c-w(L^{\ast
})+w(G_{2})-w(G_{1})$. Therefore, $G_{12}$ is a feasible solution for $S$
since the capacity of $S$ is greater than the capacity of $S_{k}^{\ast }$ by
at least $w(G_{2})-w(G_{1})$. Finally, from inequality (\ref{eq:G1G2}), the
solution value of $G_{12}$ is not smaller than $OPT_{S_{k}^{\ast }}$ and the
claim follows.
\end{proof}

Let $P^{H}(S)$ denote the solution value returned by $H^{\frac{1}{2}}$ when
applied to $S$. Then we have the following

\begin{lemma}
$p(L)+P^{H}(S)\geq (1-4\varepsilon )OPT$.
\end{lemma}

\begin{proof}
Observe that by Lemma \ref{Lemma:Z} and inequality (\ref{Eq:ineq1}), we have
\begin{equation}
OPT_{S}\geq OPT_{S^{\ast }}-2\varepsilon OPT.  \label{Eq:ineq3}
\end{equation}
We distinguish between two cases.

\begin{enumerate}
\item If \bigskip $p(L^{\ast })\geq (1-\varepsilon )OPT$ then by
inequalities (\ref{Eq:ineq}), (\ref{Eq:ineq0}), (\ref{Eq:ineq2}) and (\ref%
{Eq:ineq3}), we have
\begin{eqnarray*}
p(L)+P^{H}(S) &\geq &p(L^{\ast })-2\varepsilon OPT+\frac{1}{2}OPT_{S} \\
&\geq &(1-\varepsilon )OPT-2\varepsilon OPT=(1-3\varepsilon )OPT.
\end{eqnarray*}

\item If $p(L^{\ast })<(1-\varepsilon )OPT$ then each item profit in $%
S^{\ast }$ is smaller than $\frac{(1-\varepsilon )}{\ell }OPT$. Indeed, if $%
\lambda =\ell $ then the smallest item profit in $L^{\ast }$, and hence each
item profit in $S^{\ast }$, must be smaller than $\frac{(1-\varepsilon )}{%
\ell }OPT$ (otherwise $p(L^{\ast })\geq (1-\varepsilon )OPT$); else ($%
\lambda <\ell $) by definition of $\lambda $, there are at most $\lambda $
items with profits not smaller than $a_{1}$ and therefore, each item profit
in $S^{\ast }$, must be smaller than $a_{1}=\frac{\varepsilon }{\ell }%
P^{H}\leq $ $\frac{\varepsilon }{\ell }OPT\leq \frac{(1-\varepsilon )}{\ell }%
OPT$ (since $\varepsilon \leq 1/2$). Now, we claim that the largest profit
in $S$ is at most $\frac{(1-\varepsilon )}{\ell }OPT+\frac{\varepsilon P^{H}%
}{\ell }$. Indeed, since by definition of $h$ we have $a_{h}\leq \frac{%
(1-\varepsilon )}{\ell }OPT\leq (1-\varepsilon )\frac{2P^{H}}{\ell }\leq
(\left\lfloor \frac{2}{\varepsilon }\right\rfloor -1)\frac{\varepsilon P^{H}%
}{\ell }$, it turns out that $h\leq \left\lfloor \frac{2}{\varepsilon }%
\right\rfloor -1$, and by definition of $S_{ag}(\ell )$, we have that $%
a_{h+1}=a_{h}+\frac{\varepsilon P^{H}}{\ell }$. Therefore, for each item $j$
belonging to $S$, profit $p_{j}$ is bounded by
\begin{equation*}
p_{j}\leq \frac{\varepsilon }{\ell }P^{H}+\frac{(1-\varepsilon )}{\ell }%
OPT\leq \frac{OPT}{\ell }.
\end{equation*}%
Since $OPT_{S}-P^{H}(S)\leq \max_{j\in S}p_{j}$ (see inequality (\ref%
{Eq:ineq})), we have
\begin{eqnarray*}
p(L)+P^{H}(S)+\frac{OPT}{\ell } &\geq &p(L)+OPT_{S} \\
&\geq &p(L^{\ast })+OPT_{S^{\ast }}-4\varepsilon \cdot OPT=(1-4\varepsilon
)OPT.
\end{eqnarray*}
\end{enumerate}
\end{proof}

By the previous lemma, steps (S-1)-(S-5) return a solution that cannot be
worse than $(1-4\varepsilon )OPT$. Thus, we have proved the following

\begin{theorem}
There is an PTAS for the k-item knapsack problem requiring linear space and $%
O(n+k\cdot (1/\varepsilon )^{O(1/\varepsilon )})$ time.
\end{theorem}

To compare our algorithm with the one provided in \cite{Caprara:1998:NFP}
notice that the running time complexity of the latter is $O(n^{\lceil
1/\varepsilon \rceil -1})$, whereas our scheme is linear. As in \cite%
{Caprara:1998:NFP}, our algorithm can be easily modified to deal with the
\textit{Exact k-item Knapsack Problem}, that is a kKP in which the number of
items in a feasible solution must be exactly equal to $k$. The time and
space complexities, and the analysis of the resulting algorithm are
essentially the same as the one described above.
Compare also with the general problem solver developed in
\cite{Hutter:01fast}, where the elimination of a multiplicative
constant (here $1/\varepsilon$) led to a larger additive constant (here
$(1/\varepsilon)^{O(1/\varepsilon)}$).

\section{An improved FPTAS for kKP}

The main goal of this section is to present a different combination of
arithmetic and geometric rounding techniques. Moreover we propose an
improved fully polynomial time approximation scheme that runs in $%
O(n+k/\varepsilon ^{4}+1/\varepsilon ^{5})$ time. First we discuss
separately the different steps in details, then we state the main algorithm
and summarize the results in Section \ref{Sect:FPTASsmall}.

We start partitioning the set of items in two subsets $\mathcal{L}=\left\{
j:p_{j}>\varepsilon P^{H}\right\} $ and $\mathcal{S}=\left\{ j:p_{j}\leq
\varepsilon P^{H}\right\} $. Let us say that $\mathcal{L}$ is the set of
\emph{large} items, while $\mathcal{S}$ the set of \emph{small} items.
Observe that the number of large items in any feasible solutions is not
greater than $\lambda =\min \left\{ k,\left\lfloor 2/\varepsilon
\right\rfloor \right\} $, since $OPT\leq 2P^{H}$.

\subsection{Dynamic programming for large items}

In principle, an optimal solution could be obtained in the following way.
Enumerate all different solutions for items in $\mathcal{L}$, i.e., consider
all different sets $U\subseteq \mathcal{L}$ such that $w(U)\leq c$ and $%
|U|\leq k$. For each of these $U$, compute a set $T\subseteq \mathcal{S}$
such that $w(T)+w(U)\leq c$, $|U|+|T|\leq k$ and $p(T)$ is maximized. Select
from these solutions one with the largest overall profit. One of the
problems with this approach is that constructing all possible solutions for
items in $\mathcal{L}$ would require considering $n^{O(1/\varepsilon )}$
cases. To avoid the exponential dependence on $1/\varepsilon $ (our aim is
to obtain a fully polynomial approximation scheme), we will not treat
separately all of these solutions. We begin with the description of a basic
procedure that generates a list of all ``interesting'' feasible combinations
of profit and number of selected large items. Each such combination is
represented by a pair $(a,l)$, for which there is a subset of items $%
U\subseteq \mathcal{L}$ with $p(U)=a$, $|U|=l$ and $w(U)\leq c$. Moreover $%
w(U)$ is the smallest attainable weight for a subset of large items with
profit at least equal to $a$ and cardinality at most $l$. This list of all
``interesting'' feasible combinations is computed by using a
pseudopolynomial dynamic programming scheme. Clearly, an optimal solution
can be computed by using only the subsets $U$ of large jobs associated to
each pair $(a,l)$. The time complexity will be then reduced, with $1-O\left(
\varepsilon \right) $ loss, by applying arithmetic and geometric rounding
techniques, as described in Section \ref{Sect:serialGeoArith}.

Let $\alpha $ be the number of large items, and let $\beta $ denote the
number of all distinct feasible solution values obtained by considering only
large items, i.e. $\beta $ is the size of set
\begin{equation*}
V=\left\{ p(U)|U\subseteq \mathcal{L}\text{ and }w(U)\leq c\ \text{and }%
|U|\leq k\right\} .
\end{equation*}%
A straightforward dynamic programming recursion which has time complexity $%
O(\alpha \beta \lambda )$ and space complexity $O(\lambda ^{2}\beta )$ (see
\cite{Caprara:1998:NFP}), can be stated as follows. Let us renumber the set
of items such that the first $1,...,|L|$ items are large. Denote by function
$g_{i}(a,l)$ for $i=1,...,|L|$, $a\in V$, $l=1,...,\lambda $, the optimal
solution of the following problem:
\begin{equation*}
g_{i}(a,l)=\min \sum_{j=1}^{i}w_{j}x_{j}:\left\vert
\begin{array}{l}
\sum_{j=1}^{i}p_{j}x_{j}=a; \\
\sum_{j=1}^{i}x_{j}=l; \\
x_{j}\in \left\{ 0,1\right\} ,j=1,...,i.%
\end{array}%
\right\}
\end{equation*}%
One initially sets $g_{0}(a,l)=+\infty $ for all $l=0,...,\lambda $, $a\in V$%
, and then $g_{0}(0,0)=0$. Then, for $i=1,...,|L|$ the entries for $g_{i}$
can be computed from those of $g_{i-1}$ by using the formula
\begin{equation*}
g_{i}(a,l)=\min \left\{
\begin{array}{l}
g_{i-1}(a,l), \\
g_{i-1}(a-p_{i},l-1)+w_{i}\text{ \ \ if }l>0\text{ and }a\geq p_{i}%
\end{array}%
\right\} .
\end{equation*}

Since $\beta =O(P^{H})$ the described dynamic programming algorithm is only
pseudopolynomial. In order to reduce the time complexity, we first
preprocess large items by using a combination of arithmetic and geometric
rounding techniques, then we apply the above dynamic programming scheme. We
start analyzing the two rounding techniques separately, then we show how to
combine them.

\subsubsection{Geometric rounding}

The time complexity of the described dynamic programming can be reduced by
decreasing the number $\alpha $ of large items and the number $\beta $ of
distinct solution values.

We observed in Section \ref{Sect:reduced N} that if we want to reduce as
much as possible the number of large items it is convenient to use geometric
rounding. Consider the geometric sequence $S_{g}(\gamma )$ described in
Section \ref{Sect:Geometric rounding}. By applying the geometric rounding
technique with $\gamma =\lambda $, the number $\alpha $ of large items can
be reduced from $O(n)$ to $O(1/\varepsilon ^{2})$ with $1-\varepsilon $ loss.

The next step is to compute the number of possible solution values after
geometric rounding, i.e. the cardinality $\beta $ of set $V$ after that all
profit values of large items have been geometrically rounded. The main
result of this section is stated as follows.

\begin{theorem}
\label{Th:GeoSumBounds}The number of solution values after geometric
rounding can be exponential in ${\frac{1}{\varepsilon }}$.
\end{theorem}

By the above theorem it follows that the running time of dynamic programming
after geometric rounding is a constant that may depend exponentially on ${%
\frac{1}{\varepsilon }}$. Therefore, to avoid this exponential dependence on
${\frac{1}{\varepsilon }}$, we will look at other rounding techniques.

\paragraph{Proof of Theorem \protect\ref{Th:GeoSumBounds}.}


In the remaining part of this subsection we prove Theorem \ref%
{Th:GeoSumBounds}. The goal is to derive a lower bound on the number of
possible solution values after geometric rounding, i.e.\ a lower bound on $%
|V|$. Recall that we defined the geometric sequence $a_{i}={\frac{%
\varepsilon P^{H}}{\lambda }}({\frac{1}{1-\varepsilon }})^{i-1}$ in Section %
\ref{Sect:Geometric rounding}, and here we assume that $\gamma =\lambda $.
We focus on worst-case analysis. With this aim let us consider an instance $%
I $ that after geometric rounding has at least $\left\lfloor
P^{H}/a_{i}\right\rfloor $ items for each distinct profit value $a_{i}$.
Moreover, we assume that $w_{j}=p_{j}$ for every $j\in \mathcal{L}$, $%
c=P^{H} $ and $k\geq 1/\varepsilon $. By definition of instance $I$ we see
that every subset $U$ with $p(U)<P^{H}$ is a feasible solution. Indeed, we
have $w(U)<P^{H}=c$ and $|U|<p(U)/(\min_{j\in \mathcal{L}}p_{j})<1/%
\varepsilon \leq k$. By the previous arguments we see that $|V|$ is bounded
by below by the number of solution values $y=\sum_{i=0}^{\infty
}c_{i}a_{i+1}<P^{H}$ with $\mathbf{c}=(c_{0},c_{1},...)\in
I\!\!N_{0}^{\infty }$, where $I\!\!N_{0}^{\infty }$ is the set of sequences
with non-negative integer components. Inserting $a_{i}$ and $\varepsilon ={%
\frac{\varepsilon ^{\prime }}{1+\varepsilon ^{\prime }}}$ this is equivalent
to
\begin{equation}
\sum_{i=0}^{\infty }c_{i}\varepsilon ^{\prime }(1+\varepsilon ^{\prime
})^{i}<\lambda (1+\varepsilon ^{\prime }).  \label{cleqgam}
\end{equation}
Clearly
\begin{equation}
\sum_{i=0}^{\infty }c_{i}\varepsilon ^{\prime }(1+\varepsilon ^{\prime
})^{i}<1.  \label{cleq1}
\end{equation}
implies (\ref{cleqgam}), since $\lambda (1+\varepsilon ^{\prime })>1$.

For simplicity of notation we replace $\varepsilon ^{\prime }$ with $%
\varepsilon $, and we focus on the cardinality of sets of the form
\begin{equation*}
R_{\varepsilon }^{d}\;:=\;\left\{ y<1:y=\sum_{i=0}^{d-1}c_{i}\varepsilon
(1+\varepsilon )^{i},\quad \mathbf{c}\in I\!\!N_{0}^{d}\right\} ,
\end{equation*}
where $I\!\!N_{0}^{d}$ is the set of $d$-dimensional vectors $\mathbf{c}%
=(c_{0},c_{1},...,c_{d-1})$ with non-negative integer components. It is more
easy to find lower bounds on the set of vectors
\begin{equation}
S_{\varepsilon }^{d}\;:=\;\left\{ \mathbf{c}\in
I\!\!N_{0}^{d}:\sum_{i=0}^{d-1}c_{i}\varepsilon (1+\varepsilon
)^{i}<1\right\}  \label{Sdef}
\end{equation}
itself. To consider $|S_{\varepsilon }^{d}|$ instead of $|R_{\varepsilon
}^{d}|$ is justified by the following Lemma.

\begin{lemma}
\label{lemOnto} For rational and for transcendental $\varepsilon >0$, the
sets $R_{\varepsilon }^{d}$ and $S_{\varepsilon }^{d}$ have the same
cardinality, i.e.\ the mapping $f:S_{\varepsilon }^{d}\rightarrow
R_{\varepsilon }^{d}$ with $f(\mathbf{c})=\sum_{i=0}^{d-1}c_{i}\varepsilon
(1+\varepsilon )^{i}$ is one-to-one and onto.
\end{lemma}

\begin{proof}[\textbf{Proof for transcendental $\protect\varepsilon$}]
\begin{equation*}
y=y^{\prime }\quad \Leftrightarrow \quad \sum_{i=0}^{d-1}b_{i}x^{i}=0\quad
\text{with}\quad b_{i}:=c_{i}-c_{i}^{\prime }\in Z\!\!\!Z\quad \text{and}%
\quad x:=1+\varepsilon
\end{equation*}
A real number is said to be transcendental if it is not the root of a
polynomial with integer coefficients. For transcendental $\varepsilon $ also
$x$ is transcendental, which implies $b_{i}\equiv 0$. Hence, $y=y^{\prime }$
implies $\mathbf{c}=\mathbf{c}^{\prime }$. Obviously $\mathbf{c}=\mathbf{c}%
^{\prime }$ implies $y=y^{\prime }$. This proves that $S_{\varepsilon }^{d}$
has the same cardinality as $R_{\varepsilon }^{d}$ for transcendental $%
\varepsilon $.

\textbf{Proof for rational $\varepsilon$.} Assume by contradiction that
there are two different vectors $\mathbf{c}\neq \mathbf{c}^{\prime }\in {%
I\!\!N}_{0}^d$ with same solution value $y=y^{\prime }$. Let $n$ be the
largest index $i$ such that $c_{i}\neq c_{i}^{\prime }$. Furthermore, let $%
\varepsilon ={\frac{p}{q}}-1>0$ be rational with $p>q\in{I\!\!N} $ having no
common factors. With $\mathbf{b}:=\mathbf{c}-\mathbf{c}^{\prime } $ we have
\begin{eqnarray*}
0\; &=&\;[y-y^{\prime }]\;=\;\sum_{i=0}^{d-1}b_{i}\varepsilon(1+\varepsilon
)^{i}\; \\
&=&\;\varepsilon\sum_{i=0}^{n}b_{i}{\frac{p^{i}}{q^{i}}}\;=\;{\frac{%
\varepsilon}{q^{n}}}\Big[b_{n}p^{n}+q\overbrace{%
\sum_{i=0}^{n-1}[b_{i}p^{i}q^{n-i-1}]}^{integer}\Big]
\end{eqnarray*}
Since the last term $q\sum [...]$ is a multiple of $q$, $y-y^{\prime }$ can
only be zero if also $b_{n}p^{n}$ is a multiple of $q$. With $p$ also $p^{n}$
has no common factor with $q$, hence $b_{n}$ must itself be a multiple of $q$%
. The sum in (\ref{Sdef}) can only be less than 1 if each term is less than
1, i.e.\ $c_{i}\varepsilon (1+\varepsilon )^{i}<1$. This implies
\begin{equation}
c_{i}<\varepsilon ^{-1}(1+\varepsilon )^{-i}\leq \varepsilon ^{-1}\quad
\text{for all}\quad i.  \label{ckbnd}
\end{equation}
Together we get
\begin{equation*}
0\leq c_{n}^{_{(}}{\!\displaystyle^{\prime }}^{_{)}}<{\frac{1}{\varepsilon }}%
={\frac{q}{p-q}}<q \;\Rightarrow\; |b_{n}|=|c_{n}-c_{n}^{\prime }|<q
\;\Rightarrow\; b_{n}=0 \;\Rightarrow\; c_{n}=c_{n}^{\prime }
\end{equation*}
which contradicts our assumption $c_{n}\neq c_{n}^{\prime }$. Hence, $%
y=y^{\prime }$ implies $\mathbf{c}=\mathbf{c}^{\prime }$. Again, that $%
\mathbf{c}=\mathbf{c}^{\prime }$ implies $y=y^{\prime }$ is obvious. This
shows that $S_{\varepsilon }$ has the same cardinality as $R_{\varepsilon }$
for rational $\varepsilon $.
\end{proof}

We don't know whether Lemma \ref{lemOnto} also holds for algebraic $%
\varepsilon$. The following Lemma lower bounds $S_\varepsilon^\infty$.

\begin{lemma}
\label{lemGeoBnd} $|S_\varepsilon^\infty| \;\geq\; Ce^{B/\varepsilon} \quad%
\text{with}\quad B=0.3172... \quad\text{and}\quad C=0.3200... $
\end{lemma}

\begin{proof}
From (\ref{ckbnd}) we see that all $c_{i}$ are zero for too large $i$ ($%
c_{i}=0\,\forall i\geq d_{max}:=\lceil {\frac{\ln (1/\varepsilon )}{\ln
(1+\varepsilon )}}\rceil $). This shows that $|S_{\varepsilon }^{1}|\leq
|S_{\varepsilon }^{2}|\leq ...\leq |S_{\varepsilon
}^{d_{max}}|=|S_{\varepsilon }^{d_{max}+1}|=...=|S_{\varepsilon }^{\infty }|$%
. The main idea in the following is to relate $S_{\varepsilon }^{d}$ to the
volume of a $d$-dimensional simplex with volume larger than $%
Ce^{B/\varepsilon }$ for suitable $d$.

We define a $d$-dimensional subset $U_{\varepsilon }^{d}\subset {I\!\!R}^{d}$%
, which is the disjoint union of unit cubes $[c_{0},c_{0}+1)\times ...\times
\lbrack c_{d-1},c_{d-1}+1)$ for every $\mathbf{c}\in S_{\varepsilon }^{d}$.
This set can be represented in the following form
\begin{equation*}
U_{\varepsilon }^{d}\;:=\;\left\{ \mathbf{r}\in \lbrack 0,\infty
)^{d}:\sum_{i=0}^{d-1}\lfloor r_{i}\rfloor \varepsilon (1+\varepsilon
)^{i}<1\right\}
\end{equation*}
The volume $\text{Vol}(U_{\varepsilon }^{d})$ coincides with the cardinality
of set $S_{\varepsilon }^{d}$ since each point in $S_{\varepsilon }^{d}$
corresponds to exactly one unit cube in $U_{\varepsilon }^{d}$, each having
volume 1. Furthermore, let us define the $d$-dimensional (irregular)
tetrahedron
\begin{equation*}
T_{\varepsilon }^{d}\;:=\;\left\{ \mathbf{r}\in \lbrack 0,\infty
)^{d}:\sum_{i=0}^{d-1}r_{i}\varepsilon (1+\varepsilon )^{i}<1\right\}
\end{equation*}
Obviously $T_{\varepsilon }^{d}\subseteq U_{\varepsilon }^{d}$, since $%
\lfloor r_{i}\rfloor \leq r_{i}$. So we have $|S_{\varepsilon }^{\infty
}|\geq |S_{\varepsilon }^{d}|=\text{Vol}(U_{\varepsilon }^{d})\geq \text{Vol}%
(T_{\varepsilon }^{d}).$ The tetrahedron $T_{\varepsilon }^{d}$ is
orthogonal at the vertex $\mathbf{r}=0$. The edges $\mathbf{r}%
=(0,...,0,r_{i},0,...,0)$ have lengths $[\varepsilon (1+\varepsilon
)^{i}]^{-1}$, $i=0...d-1$. Hence, the volume of the tetrahedron is
\begin{eqnarray*}
\text{Vol}(T_{\varepsilon }^{d})\; &=&\;{\frac{1}{d!}}\prod_{i=0}^{d-1}[%
\varepsilon (1+\varepsilon )^{i}]^{-1}\;=\;[d!\,\varepsilon
^{d}(1+\varepsilon )^{d(d-1)/2}]^{-1}\; \\
&\geq &\;e^{[-(d\varepsilon )\ln (d\varepsilon )-{\frac{1}{2}}(d\varepsilon
)^{2}]/\varepsilon }\;=\;e^{f(d\varepsilon )/\varepsilon }
\end{eqnarray*}
with $f(x):=-x\,\ln x-{{\frac{1}{2}}}x^{2}$. In the inequality we replaced $%
d-1$ by $d$ and used $d!\leq d^{d}$ and $1+\varepsilon \leq e^{\varepsilon }$%
. The best bound is found by maximizing $e^{f(\varepsilon d)/\varepsilon }$
w.r.t.\ $d$, or equivalently by maximizing $f(x)$ w.r.t.\ $x$. We have $%
-f^{\prime }(A)=\ln A+1+A=0$ for $A=0.2784...$. Hence, $d$ should be chosen
as ${\frac{A}{\varepsilon }}$, but since $d$ is integer we have to round
somehow, for instance $d=\lfloor {\frac{A}{\varepsilon }}\rfloor $. Note
that $|S_{\varepsilon }^{d}|$ increases with $d$, but our approximation
becomes crude for $d$ near $d_{max}$. This is the reason why the maximizing $%
d$ is less than $d_{max}$. For small $\varepsilon $ we have $f(\varepsilon
\lfloor {\frac{A}{\varepsilon }}\rfloor )\approx f(\varepsilon {\frac{A}{%
\varepsilon }})=f(A)=-A\,\ln A-{{\frac{1}{2}}}A^{2}=A(1+{{\frac{1}{2}}}%
A)=:B=0.3172...$ with corrections of order $O(\varepsilon )$. This
establishes an asymptotic bound $\sim e^{B/\varepsilon }$. More exactly, one
can show that $f(\varepsilon \lfloor {\frac{A}{\varepsilon }}\rfloor )\geq
f(A)(1-{\frac{\varepsilon }{A}})$ for all $\varepsilon $. This yields the
bound
\begin{equation*}
|S_{\varepsilon }^{\infty }|\;\geq \;\max_{d}\text{Vol}(T_{\varepsilon
}^{d})\;\geq \;e^{f(A)(1-{\frac{\varepsilon }{A}})/\varepsilon
}\;=\;Ce^{B/\varepsilon }\quad \text{with}\quad C=e^{-B/A}=0.3200...
\end{equation*}
\end{proof}

The coefficient $B$ can be improved to $0.7279...$ for sufficiently small $%
\varepsilon$ by using the more accurate Stirling approximation for $d!$.

Using Lemma \ref{lemOnto} and \ref{lemGeoBnd} it is now easy to lower bound
the number of possible solution values for geometric profit distribution.
>From Lemma \ref{lemGeoBnd} we know that (\ref{cleq1}) has at least $%
Ce^{B/\varepsilon }$ solution vectors $\mathbf{c}$ and from Lemma \ref%
{lemOnto} that (\ref{cleqgam}) has at least $Ce^{B/\varepsilon }$ solution
values $y$ for rational $\varepsilon $, and the proof of Theorem \ref%
{Th:GeoSumBounds} follows.

\subsubsection{Arithmetic rounding}

Alternatively, we may think to apply arithmetic rounding to the set of large
items. Let us consider the arithmetic sequence $S_{a}(\gamma )$ described in
Section \ref{Sect:arithmetic rounding}. By applying the arithmetic rounding
technique with $\gamma =\lambda $, we observe the number of large items can
be reduced to be bounded by $O(\frac{1}{\varepsilon ^{2}}\ln \frac{1}{%
\varepsilon })$ with $1-\varepsilon $ loss. Moreover each element of set $V$
is equal to $\frac{\varepsilon P^{H}}{\lambda }i$ for some $i=\lambda
,\lambda +1,...,2\left\lfloor \lambda /\varepsilon \right\rfloor $. It
follows that the size of set $V$ is bounded by $O(1/\varepsilon ^{2})$, and
the overall time of the dynamic programming algorithm is now $O(\frac{1}{%
\varepsilon ^{5}}\ln \frac{1}{\varepsilon })$. We see that in comparison to
the geometric rounding and although the number of large items is larger, the
arithmetic rounding technique is able to reduce much more the size of set $V$%
. However and again, we can take advantage from both techniques by combining
them as described in the following.

\subsubsection{Serial Geometric \& Arithmetic rounding\label%
{Sect:serialGeoArith}}

We first apply geometric rounding with $1-\varepsilon $ loss. This reduces
the number of large items to be bounded by $O(1/\varepsilon ^{2})$. Then,
with $1-\varepsilon $ loss, we apply arithmetic rounding on the reduced set
of large items. Clearly the latter does not increase the number of items and
each profit value is now equal to $\frac{\varepsilon P^{H}}{\lambda }i$ for
some $i=\lambda ,\lambda +1,...,2\left\lfloor \lambda /\varepsilon
\right\rfloor $. By using this set of items with profits rounded by using
geometric first and arithmetic rounding then, the size of set $V$ has a
bound of $O(1/\varepsilon ^{2})$, and the overall time of the dynamic
programming algorithm is $O(1/\varepsilon ^{5})$. We call this combination a
\textit{Serial Geometric \& Arithmetic rounding technique}.

\subsection{Adding small items\label{Sect:FPTASsmall}}

In the following we show how to add the small items. First, with $%
1-2\varepsilon $ loss, we reduce the number of small items to be $%
O(k/\varepsilon )$ by using the Parallel Arithmetic \& Geometric rounding
(see Section \ref{Sect:arith_geo}). Then, for each pair $(a,l)$ in the final
list, fill in the remaining knapsack capacity $c-g_{|L|}(a,l)$ with at most $%
k-l$ small items, by using algorithm $H^{\frac{1}{2}}$ for kKP \cite%
{Caprara:1998:NFP}. These small items yield total profit $%
P^{H}(c-g_{|L|}(a,l),k-l)$. By inequality (\ref{Eq:ineq}) and by definition
of small items, we have
\begin{equation}
P^{H}(c-g_{|L|}(a,l),k-l)+\varepsilon P^{H}\geq OPT(c-g_{|L|}(a,l),k-l),
\label{eq:ineqFPTAS}
\end{equation}
where $OPT(c-g_{|L|}(a,l),k-l)$ is the optimal solution value obtained by
using at most $k-l$ small items and knapsack capacity $c-g_{|L|}(a,l)$. The
approximate solution, a combination of large and small items, is chosen to
yield profit $P$, where
\begin{equation*}
P=\max_{(a,l)}\left\{ a+P^{H}(c-g_{|L|}(a,l),k-l)\right\}
\end{equation*}
By inequality (\ref{eq:ineqFPTAS}) and since our algorithms considers all
the ``interesting'' pairs $(a,l)$ with $1-O(\varepsilon )$ loss, it is easy
to verify that $P$ is $1-O(\varepsilon )$ times the optimal solution.

To summarize, the steps of the FPTAS are as follows.

\begin{enumerate}
\item[(S-1)] Partition the set of items into ``large'' and ``small''. Apply
the Serial Geometric \& Arithmetic rounding technique to the set of large
items. Apply the Parallel Arithmetic \& Geometric rounding technique to the
set of small items.

\item[(S-2)] Solve for the ``large'' items using dynamic programming:
generate a list of all ``interesting'' feasible combinations $(a,l)$ of
profit $a$ and number $l$ of selected large items.

\item[(S-3)] For each pair $(a,l)$ in the final list, fill in the knapsack
by applying algorithm $H^{\frac{1}{2}}$ with the reduced set small items.

\item[(S-4)] Return the best found solution.
\end{enumerate}

Step (S-1) can be performed in $O(n)$ time. Step (S-2) takes $%
O(1/\varepsilon ^{5})$ time. Algorithm $H^{\frac{1}{2}}$ applied to the
reduced set of small items runs in $O(k/\varepsilon )$ time \cite%
{Caprara:1998:NFP}. In step (S-3) the algorithm considers $O(1/\varepsilon
^{3})$ pairs, for each one performing operations that require $%
O(k/\varepsilon )$ time. It follows that the overall running time of the
algorithm is $O(n+k/\varepsilon ^{4}+1/\varepsilon ^{5})$. The space
complexity has a bound of $O(n+1/\varepsilon ^{4})$, since the space
required by the dynamic programming is $O(\lambda ^{2}\beta )$ where $%
\lambda =O(1/\varepsilon )$ and $\beta =O(1/\varepsilon ^{2})$.

\begin{theorem}
There is a fully polynomial time approximation scheme for the k-item
knapsack problem requiring $O(n+k/\varepsilon ^{4}+1/\varepsilon ^{5})$ time
and $O(n+1/\varepsilon ^{4})$ space.
\end{theorem}

\paragraph{Acknowledgments.}

Thanks are due to Klaus Jansen for introducing us to the k-item Knapsack
Problem. We are grateful to the referees who pointed out some mistakes in
the early version of this paper.

\end{document}